\def\ba{\begin{eqnarray}}
\def\ea{\end{eqnarray}}
\def\lb{\label}
\def\nn{\nonumber \\}
\def\bi{\bibitem}
\def\g{\gamma}

\def\rr{\rightarrow}

\documentclass[12pt]{article}
\usepackage{graphicx}
 \begin{document}
\title{Multibin long-range correlations}

\author{A.Bialas and K.Zalewski \\ H.Niewodniczanski Institute of
Nuclear Physics\\ Polish Academy of Sciences\thanks{Address: Radzikowskiego
152, 31-342 Krakow, Poland}\\and\\ M.Smoluchowski Institute of Physics
\\Jagellonian University\thanks{Address: Reymonta 4, 30-059 Krakow, Poland;
e-mail:bialas@th.if.uj.edu.pl;}}
 \maketitle


keywords: long-range correlations, particle production

\begin{abstract}

A new method to study the long-range correlations in multiparticle production
is developped. It is proposed to measure the joint factorial moments or
cumulants of multiplicity distribution in {\it several} (more than two) bins.
It is shown that this step dramatically increases the discriminative power of
data.

\end{abstract}

\section{Introduction}

It is now widely recognized that long-range correlations (LRC) in rapidity give
information about the early stages of the collision. Indeed, such correlations
cannot appear at late stages in the evolution of the produced system when
longitudinal expansion separated the particles by large distances. Just after
the collision, however, the system is small enough for the  correlations to
extend through the whole system.

A special case of LRC are forward-backward correlations where one compares
particle distributions in two intervals located symmetrically in the forward
and backward hemispheres. They were extensively studied since early times of
high-energy physics \cite{kitw}.

One of the interesting issues in particle production is the question if the
produced particles "remember" the colliding projectiles, their energies,
momenta and quantum numbers. Obviously the answer depends on the kinematic
region we are considering. Close to the fragmentation region, the influence of
the projectile on the produced particle spectrum is naturally expected. In the
central rapidity region, far from the projectile fragmentation, the question
remains open. On the theoretical side there is no consensus and various models
give different answers.

An excellent review of models can be found in \cite{kitw} (see also,
\cite{bzbf}), therefore here we only quote some examples. With respect to the
question of the number and structure of particle sources, they may be divided
into three categories. In the first one, originating from the famous Landau and
Feynman papers \cite{landau,feynman}, particles produced in the central
rapidity region are decoupled from the projectiles. Thus the source of
particles is symmetric with respect to $y=0$. In the second class, like the
wounded nucleon model \cite{bbc}, particles are produced by quasi-independent
emisssion from the two colliding objects. In this case particles in the central
region come from two sources,  naturally asymmetric ones \cite{bc,bb}. There is
of course also a third class which combines the two pictures, a typical example
being the dual-parton model \cite{dpm}.

These various mechanisms can be tested (and verified) by studying the
forward-backward correlations. The essential point is that correlations for one
symmetric source are generally much stronger than those induced by two
asymmetric ones \cite{bzbf,bzdakx}. Following this general idea we recently
proposed a systematic method of investigation of the forward-backward
correlations in {\it symmetric} hadronic and heavy ion collisions \cite{bzbf}.
It was shown that such investigations allow to verify how many indepedent
sources of particles contribute to the observed distributions.

In the present paper we generalize these results in two respects:

(i) we abandon the requirement of symmetry and consider the general case of
asymmetric processes and thus also asymmetric sources;

(ii) We suggest to measure and compare particle distributions in more than two
intervals\footnote{Measurements in three intervals were  considered in
\cite{star} and \cite{lappi}.}.

This generalization of the problem allows to undertake a general discussion of
LRC and thus extends its application to other processes, like, e.g.
lepton-nucleon, hadron-nucleus and asymmetric nucleus-nucleus collisions.

We consider measurements of multiplicity moments in $B$ intervals. Following
our previous paper \cite{bzbf}, we assume that particles are produced by
indepedent sources and that the population of particles in $B$ bins from one
source is random, i.e. it is described by the multinomial
distribution\footnote{This assumption which may be understood as the definition
of a "source" is accepted in practically all published discussions of the
forward-backward correlations, see e.g. \cite{cy}. For an extensive list of
references, see \cite{kitw}.}.

We then evaluate the number of measurable (factorial) moments of the
distribution and compare it with the number of parameters in the system. This
allows to draw our main conclusion: the discriminating power of the method
increases dramatically with increasing number of intervals in which the
measurements are performed.

In the next section we present the mathematical structure of our approach. In
Section 3 the number of possible measurements and number of independent
parameters are evaluated for the general case of particle production from
independent sources and a measurement in $B$ intervals. An important special
case, when the number of independent moments is reduced by symmetry, is
discussed in the Appendix. In Section 4 the explicit formulae giving the
factorial cumulants for two models with fixed numbers of sources are derived
and commented upon. The general formulae and two examples of models with
fluctuating numbers of sources are discussed in Section 5. The summary of the
results can be found in the last section.

\section{Formulation of the problem}

Following the assumptions explained in Introduction, we write the
 generating function for the particle
distributions in the $B$ bins  in the form

\begin{equation}\label{phi0}
\Phi(z_1,...,z_B)=\left\langle\prod_{i=1}^N
\phi_i^{w_i}(p_{1i}z_1+...+p_{Bi}z_B)\right\rangle,
\end{equation}
where $\phi_i$ is the generating function for the $i$-th source, $p_{ki}$ is
the probability that the $i$-th source sends particles into the $k$-th bin and
$w_i$ is the number of sources of type $i$. The angular brackets denote
averaging over the multiplicities $w_i$ (they can be omitted if the $w_i$ do
not fluctuate). Since the generating functions $\phi_i$ do not have to be all
different, one can assume without loss of generality that each of the numbers
$w_i$ can only take the value zero or one (Section 5.1), but sometimes it is
more convenient to assume that $w_i$ can be any nonnegative integer (Section
5.2). The measurable (factorial) moments are given by

\begin{equation}\label{momgen} F_{i_1,\ldots,i_B}
\equiv \left\langle\prod_{j=1}^B \frac{n_j!}{(n_j-i_j)!}\right\rangle =
\frac{\partial^r\Phi(z_1,\ldots,z_B)} {\partial z_1^{i_1}\ldots\partial
z_B^{i_B}}.
\end{equation}
where $n_j$ is the number of particles in bin $j$. Here and henceforth all
derivatives are taken at $z=z_1=...z_B=1$. Note that

\begin{equation}\label{Fri}
  \frac{d^r\phi_n(z)}{dz^r} \equiv F^{(r)}_n
\end{equation}
is the r-th factorial moment of the distribution of the total number
 of particles sent by source $n$ to all  bins.

If the numbers and nature of the sources do not fluctuate, it is advantageous
to introduce the cumulants

\begin{equation}
  f_{i_1\ldots i_B} =  \frac{\partial^r\log \Phi(z_1,\ldots,z_B)}
{\partial z_1^{i_1}\ldots\partial z_B^{i_B}}
\end{equation}
which, as is easily derived from (\ref{phi0}), can be expressed as

\begin{equation}\label{cumula}
 f_{i_1\ldots i_B} =
\sum_{n=1}^Np_{1n}^{i_1}\ldots p_{Bn}^{i_B}f_n^{(r)},
\end{equation}
where

\begin{equation}\label{fixed}
  f_n^{(r)} = \frac{d\log \phi_n(z)}{dz^r}
\end{equation}
are cumulants of the distribution produced by the $n$-th source.

Let us  also note here that using (\ref{cumula}) and the identity
\begin{equation}\label{iden}
\sum \frac{r!}{j_1!...j_B!}p_1^{j_1}...p_B^{j_B}=(p_1+...+p_B)^r=1,
\end{equation}
one finds the very useful  relation

\begin{equation}\label{sumrul}
\sum\frac{r!}{j_1!...j_B!}f_{j_1,\ldots,j_B} = \sum_{n=1}^Nf^{(r)}_n.
\end{equation}

\section{Counting of parameters}

Consider a general situation of $N$ groups of independent sources, all sources
in one group being identical, and $B$ bins. No symmetry relations among groups
are assumed.

We first evaluate the number of moments which can be measured. To this end we
observe that each moment has $B$ indices: $F_{i_1i_2...i_B}$. Define the rank
$r$ of the moment as \ba r=i_1+i_2+...+i_B. \ea

The number of moments at given $r$ and $B$, m(r,B), is the solution of the
well-known combinatorial problem: in how many ways can one distribute $r$
identical objects among $B$ boxes:

\begin{equation}\label{nummom}
 m(r,B) = \frac{(r+B-1)!}{r!(B-1)!}
\;\; \rr \;\; \sum_{r=1}^{r_{max}}m(r,B)=\frac{(B+r_{max})!} {B!r_{max}!}-1
\end{equation}

The next thing we want to know is the number of parameters in the model. First,
there are $N(B-1)$ independent probabilities.  In addition we need also, for
each kind source, the derivatives of order up to $r_{max}$ of the multiplicity
generating function $[\phi_n(z)]^{w_n}$:

\begin{equation}\label{ftilde}
\tilde{F}_n^{(r)} = \frac{d^r}{dz^r}\left[\phi_n^{w_n}(z)\right].
\end{equation}
They are polynomials in the random variable $w_n$.

The expressions for the measurable moments of order $r$ contain the averages

\begin{equation}\label{effe}
  \left\langle \tilde{F}_1^{(r_1)}\ldots
 \tilde{F}_N^{(r_N)}\right\rangle,\qquad \sum_{n=1}^Nr_n = r.
\end{equation}

When the multiplicity distribution for sources is not known, each of these
averages is an independent parameter. Using the same combinatorial formulas as
before, we thus find that the number of independent parameters is

\begin{equation}
P(B,N,r_{max})=N(B-1)+\frac{(r_{max}+N)!}{r_{max}!N!} -1. \lb{no}
\end{equation}

When the distribution of numbers of  sources $W(w_1,\ldots,w_N)$ is known, all
averages (\ref{effe}) are  determined in terms of $F_n^{(r_n)}$ and therefore
the number of independent parameters is \ba P(B,N,r_{max})=N(B+r_{max}-1).
\lb{yes} \ea

Thus we finally obtain for the number of parameter-independent constraints
between the measurable quantities

\ba
 C(N;B;r_{max})= \frac{(B+r_{max})!} {B!r_{max}!}-1 -P(B,N,r_{max})
\ea where $P(B,N,r_{max})$ is given by (\ref{no}) or  (\ref{yes}).

To obtain tests, we demand that $C\geq 1$. It is clear that for any $N$ and
$r_{max} \geq 2$ one can always find $B$ such that this condition is satisfied.

For practical reasons, one has to keep  $r_{max}$ rather small,
 say $2$ or $3$. In Table 1 we give the minimal number of bins necessary
to obtain parameter-independent constraints.

Table 1

 \vspace{0.5cm}

\begin{tabular}{|c|p{2.6cm}|p{2.6cm}|p{2.6cm}|p{2.6cm}|}
  \hline
  \multicolumn{5}{|c|}{Minimal numbers of bins necessary to get predictions}\\
  \hline\hline
   &\multicolumn{2}{c|}{Fixed number of sources}&\multicolumn{2}{c|}{Fluctuating number of sources}\\ \hline
  $N$ & $r_{max}=2$ & $r_{max}=3$ & $r_{max}=2$ & $r_{max}=3$ \\ \hline
  1 & 2 & 2 & 2 & 2 \\
  2 & 3 & 2 & 4 & 3 \\
  3 & 5 & 3 & 6 & 4 \\ \hline\hline
\end{tabular}
\vspace{0.5cm}

To illustrate possible applications of this general discussion, we present in
the next two sections four examples of specific models of particle production
which can be tested in this way.

\section{ Fixed number of sources}

For a fixed number of sources the measurable cumulants are given by
(\ref{cumula}). Below we give two specific examples.

\subsection{Landau model: one source}

In the Landau model there is just one source of particles, resulting from
hydrodynamic expansion of the remnant of the two projectiles (just after
collision the remnant is concentrated at $y_{cm}=0$). For one source ($N=1$)
already at $r_{max} = 1$ the number of measurable moments is equal to the
number of parameters. Therefore, it is possible to determine all the
probabilities $p_j$ from the moments (cumulants) of rank one.  For each $r>1$
there is one more parameter, $f^{(r)}=d^r[\log\phi(z)]/dz^r$. Using
(\ref{cumula}), this parameter can be evaluated from any measured moment of
rank $r$. Indeed, for one source we simply have
\begin{equation}
f^{(r)}_{j_1,\ldots,j_B} = p_1^{j_1}...p_B^{j_B} f^{(r)}
\end{equation}
where the subscript denoting the source was dropped. Since all probabilities
are already determined from the moments of rank one, this formula allows to
evaluate $f^{(r)}$ and thus all other measurable cumulants of rank $r$.

\subsection{Deep inelastic scattering: two sources}

In deep inelastic scattering there are at least two different sources: the
proton and photon remnants.  It is thus interesting to investigate if these two
sources are sufficient to describe the data. In this section we show that the
hypothesis of two sources gives indeed strong constraints on particle
correlations.

Following the argument of Section 3, we consider $B$ bins located anywhere
along the direction of the incident photon. We thus have $2(B + r_{max} - 1)$
parameters (the probabilities $p_{j\g}$, $p_{jP}$ and the cumulants
$f^{(r)}_\g$, $f^{(r)}_P$). Let us denote by $f^{(r)}_j$ the measurable
cumulant of order $r$ of the distribution of particles in the bin $j$. We show
below how, using the measured cumulants $f_j^{(r)}$ for $r\leq 2$, one can
determine all the probabilities.

Since the sum rule (\ref{sumrul}) allows to determine the sum of the cumulants
$f^{(r)}_+=f^{(r)}_\g+f^{(r)}_P$ for any $r$, we are left with with $r$ free
parameters $f^{(r)}_-=f^{(r)}_\g-f^{(r)}_P$ which should be sufficient to
predict the correct values of the other measured cumulants. Instead of the
parameters $p_{j\g}$ $p_{jP}$ it is more convenient to use

\begin{equation}
  p_{j\pm} = p_{j\g} \pm p_{jP}.
\end{equation}

 As already mentioned, for each $r$ the parameter $f_+^{(r)}$ can be
obtained directly from formula (\ref{sumrul}). For $r=1$ and $r=2$ we have \ba
f_+^{(1)}=\sum_{i=1}^B<n_i>\;;\;\;\; f_+^{(2)}=\sum_{i=1}^B<n_i(n_i-1)> +
2\sum_{i>j}^B <n_in_j> \ea where $n_i$ is the number of particles observed in
bin $i$.

Let us consider first the cumulants of order one (they coincide with the
moments of order one, i.e. average multiplicities). From formula (\ref{cumula})
one gets
\begin{equation}\label{pjp}
  f^{(1)}_j = \frac{1}{2}\left(p_{j+}f_+^{(1)} + p_{j-}f_-^{(1)} \right)
\;\rr\;  p_{j+} = \frac{2f^{(1)}_j - p_{j-}f_-^{(1)}}{f_+^{(1)}},
\end{equation}
which together with the sum rule for $f_+^{(1)}$ eliminates $B$ parameters.

Let us consider now the cumulants $ f^{(2)}_j$. From (\ref{cumula}) we have \ba
4 f^{(2)}_j  = p_{j-}^2f_+^{(2)}+2p_{j-}p_{j+} f_-^{(2)}+p_{j+}^2f_+^{(2)}
\lb{pjm} \ea Using  (\ref{pjp}) to eliminate $p_{j+}$ we get a quadratic
equation for $p_{j-}$. The two solution of this equation depend on the
parameters $f_-^{(1)}$ and $f_-^{(2)}$. Thus we get $2^B$ possible sets
$p_{1-}\ldots p_{B-}$. Hopefully most of them can be eliminated by the obvious
requirement that each $p_{j-}$ must be real and that the following constraints
must be satisfied.
\begin{equation}
  |p_{j-}| < p_{j+},\qquad |p_{j-}|<1,\qquad \sum_{j=1}^Bp_{j-} = 0.
\end{equation}
Thus, if the model is consistent with  data, i.e. solutions exist, all the
probabilities are determined, though some ambiguities may be left.

For $r_{max}=2$ we have, in addition, $\frac{1}{2}B(B-1)$ cumulants of the type
$f_{110\ldots 0}$ which should be fitted with two parameters $f_-^{(1)}$ and
$f_-^{(2)}$. Increasing $r_{max}$ by one, introduces two new parameters
$f_+^{(3)}$ and $f_-^{(3)}$. The former, however, is fixed by the sum rule
(\ref{sumrul}) so that there are $m(3,B)$ new cumulants, constrained by the sum
rule which has already been used, to be fitted with one free parameter.

\section{ Fluctuating number of sources}

When the number and nature of sources fluctuate, the discussion of LRC becomes
rather involved. The reason is that the formulae expressing the measurable
moments in terms of the parameters of the model become complicated, as can be
seen later in this section. In most models, however, the sources are not
entirely arbitrary and thus these relations can be simplified.

We start with the general formulae for arbitrary number and nature of sources
and then discuss two examples, suggested respectively by  the dual parton model
and by the wounded constituent model.

\subsection{General formulae}

Let us consider the generating function (\ref{phi0}) with each $w_i$ equal zero
or one. Then

\begin{equation} w_i(w_i-1)=0. \end{equation}

This greatly simplifies the differentiations. In fact

\begin{equation}\frac{d^r}{dz^r}\phi_i^{w_i}(z) =
w_i\frac{d^r}{dz^r}\phi_i(z) = w_iF^{(r)}_i.
 \end{equation}

Below we give the formulae for the measurable factorial moments of rank 1, 2
and 3. They are written assuming that only the first (for $r=1)$, the first two
(for $r=2$), or the first three (for $r=3$) bin indices are non-vanishing.
Analogous formulae are of course valid for any other selection of bins, pairs
of bins and triplets of bins.

\begin{eqnarray}
  F_{10\ldots} = \sum_{i=1}^N\langle w_i\rangle p_{1i}F^{(1)}_i,\nonumber\\
  F_{110\ldots} = \sum_{i=1}^N\langle w_i\rangle p_{1i}p_{2i}F^{(2)}_i +
  \sum_{i\neq j}^N\langle w_iw_j\rangle p_{1i}p_{2j}F^{(1)}_iF^{(1)}_j\\
  F_{1110\ldots} = \sum_{i=1}^N\langle w_i\rangle p_{1i}p_{2i}p_{3i}F^{(3)}_i
  + \sum_{i\neq j}^N\langle w_iw_j\rangle p_{1i}p_{2i}p_{3j}F^{(2)}_iF^{(1)}_j
+\nonumber\\
\sum_{i\neq j\neq k \neq i}^N\langle w_iw_jw_k\rangle
p_{1i}p_{2j}p_{3k}F^{(1)}_iF^{(1)}_jF^{(1)}_k,\nonumber
\end{eqnarray}
where $F_i^{(r)}$ is the $r$-th factorial moment of the distribution
 of particle from source $i$ (c.f. (\ref{Fri})).  When
some indices coincide, it is enough to change correspondingly
 the bin indices of the probabilities $p$. For instance,
\ba F_{20...}=\sum_{i=1}^N\langle w_i\rangle p_{1i}^2F^{(2)}_i +
  \sum_{i\neq j}^N\langle w_iw_j\rangle p_{1i}p_{1j}F^{(1)}_iF^{(1)}_j.
\ea

\subsection{Dual parton model}

For a general nucleus-nucleus collision we have a certain number $N_L$ of
identical sources moving left, a number $N_R$ of identical sources moving right
and $N_C$ identical symmetric sources. These numbers fluctuate from event to
event and their distribution depends also on the centrality of the collision.
The left and right moving sources are mirror images of each other with respect
to cm rapidity.

We consider the case where the bins are also selected to be symmetric with
respect to $y_{cm}=0$. Then if $\phi_a(p_{1a}z_1+\ldots +p_{Ba}z_B)$, where $a$
stands for asymmetric, is the generating function for the multiplicity
distributions in the bins $1,\ldots,B$ of the particles originating from a left
moving source, then $\phi_a(p_{Ba}z_1+\ldots +p_{1a}z_B)$ is the corresponding
generating function for the particles originating from a right moving source.

Let us denote by $w_L$, $w_R$, $w_C$ the numbers of left moving, right moving
and central sources. In \cite{bzbf} we discussed mostly the case of two bins
and fixed $w_L = w_R$ and $w_C$. Here we assume an arbitrary number of bins and
a general joint probability distribution $W(w_L,w_R,w_C)$ which, however, can
be evaluated, e.g. by the Glauber method (the result will, naturally, depend on
the model adopted for particle production). Then the overall generating
function for the multiplicity distributions in the $B$ bins is

\begin{eqnarray}\label{phiw}
\Phi(z_1,...,z_B)=\sum_{w_L,w_R,w_C}W(w_L,w_R,w_C) \nonumber \\
 \left[\phi_a(p_{1a}z_{1}+...+p_{Ba}z_B)\right]^{w_L}
\left[\phi_a(p_{Ba}z_{1}+...+p_{1a}z_B)\right]^{w_R}\nn
\left[\phi_C(p_{1C}z_1+...+p_{BC}z_B)\right]^{w_C}.
\end{eqnarray}

We will denote the probabilities by $p_{iA}$ where $A = L,R,C$. Although the
probabilities $p_{iR}$ can be expressed by the probabilities $p_{iL}$, this
redundancy in the notation makes the following formulae much shorter.
Similarly, the derivatives (\ref{ftilde}) are denoted by $\tilde F_A^{(r)}$.
Using this notation, the explicit expressions for the measurable factorial
moments, obtained by differentiation of (\ref{phiw}), read

\ba
  F^{(1)}_i = \sum_A p_{iA}\langle \tilde F_A^{(1)}\rangle ;\nn
F^{(2)}_{ij} = \sum_A p_{iA}p_{jA}\langle \tilde F_A^{(2)}\rangle
 +\sum_{A \neq B}p_{iA}p_{jB}\langle
\tilde F_A^{(1)}\tilde F_B^{(1)}\rangle; \nn
  F^{(3)}_{ijk} = \sum_A p_{iA}p_{jA}p_{kA}\langle \tilde F_A^{(3)}\rangle
   + \nn+
3\sum_{A\neq B}[p_{iA}p_{jB}p_{kB} + p_{jA}p_{kB}p_{iB}
+p_{kA}p_{iB}p_{jB}]\langle \tilde F_A^{(1)} \tilde F_B^{(2)} \rangle  +
  \nonumber \\ + \sum_{A \neq B \neq C \neq A}p_{iA}p_{jB}p_{kC}\langle
\tilde F_A^{(1)}\tilde F_B^{(1)}\tilde F_C^{(1)}  \rangle. \lb{m1}
\end{eqnarray}

The parameters of the model are the $2(B-1)$ probabilities and the averages
(\ref{effe}). Their number is given by (\ref{no}) or (\ref{yes}) with $N=2$.

Since  formulae (\ref{m1}) are rather complicated, it seems that in absence of
other constraints, the best way to proceed is to try to fit them by minimizing
the $\chi^2$. If the fit works, the resulting values of the probabilities and
of the factorial moments give information about the properties of the sources.

For $pp$ scattering $w_L=w_R=1$ and thus the relations are much simpler. As
they may be easily obtained from (\ref{m1}), we discuss here only the number of
parameter-indepedent constraints. The number of possible measurements is given
the the Appendix. The number of parameters is $B+B/2-2+2r_{max}$ for $B$ even
and $B+(B+1)/2 -2+2r_{max}$ for $B$ odd. One can see that for $r_{max}=3$ there
are already 3 constraints for $B=3$ and 8 constraints for $B=4$. If one wants
to restrict the measurements to $r_{max}=2$, it is necessary to measure
distributions in at least  $5$ bins. Then one  obtains 2 constraints.

\subsection{Wounded constituent model}

In the wounded constituent model, particles are emitted by the wounded
constituents moving left or right, thus there are no central sources. The
relevant formulae can be obtained from (\ref{m1}) by putting
$\tilde{F}_C^{(r)}=0$.

They can be written in the form: \ba F^{(1)}_i = p_{iL}\langle \tilde
F_L^{(1)}\rangle+ p_{iR}\langle \tilde F_R^{(1)}\rangle ;\nn F^{(2)}_{ij} =
p_{iL}p_{jL}\langle \tilde F_L^{(2)}\rangle + p_{iR}p_{jR}\langle \tilde
F_R^{(2)}\rangle + (p_{iL}p_{jR}+p_{iR}p_{jL})\langle \tilde F_L^{(1)}\tilde
F_R^{(1)}\rangle ; \nn F^{(3)}_{ijk} =\left\{ p_{iL}p_{jL}p_{kL}\langle \tilde
F_L^{(3)}\rangle +(L\rr R)\right\} + \nn+ \left\{3[p_{iL}p_{jR}p_{kR} +
p_{jL}p_{kR}p_{iR} +p_{kL}p_{iR}p_{jR}]\langle \tilde F_L^{(1)} \tilde
F_R^{(2)} \rangle+(L\leftrightarrow R)\right\}. \lb{m2} \end{eqnarray}

The consequences  of these formulae are different for symmetric (e.g. $Au-Au$
collision) and asymmetric (e.g. $d-Au$ collision) processes.

When the probability distribution $W(w_L,w_R)$ is known, the number of
parameters at a given $r_{max}$ is $B +r_{max} - 1$. For asymmetric processes
the number of possible measurements  is given by (\ref{nummom}) and for
symmetric processes the relevant formulae are given in the Appendix. Already
for $B=3$ and $r_{max}=2$ one obtains 2 parameter-indepedent constraints for
symmetric and 5  for asymmetric processes.

When $W(w_L,w_R)$ is not known, the various averages of the moments $\langle
\tilde F^{(s)}_L \tilde F^{(r-s)}_R\rangle$, ($s=0,...,r$), have to be fitted
from data at every $r \leq r_{max}$. At given $r$ the number of independent
averages is $r+1$ for the asymmetric case and for the symmetric case it is
$r/2+1$ for $r$ even and $(r+1)/2$ for $r$ odd. It is remarkable that already
at $B=3$  parameter-indepedent constraints exist. For symmetric processes one
obtains 1 constraint for $r_{max}=2$ and 5 constaints for $r_{max}=3$. For
asymmetric processes the correponding numbers are 2 and 8.

\section{Summary }

Extending the ideas formulated in \cite{bzbf} (see also \cite{bzdakx}), a new
method to study the long-range correlations in particle production is
developed. The new proposition is to measure the factorial moments and/or
cumulants in {\it several} bins, as opposed to previous studies which were
mostly restricted to just two bins (see, however, footnote 1). It was shown
that increasing the number of bins magnifies dramatically the possibility of
discriminating between various models of particle production.

The discriminative power of the method was analyzed in the most general way.
Apart from this general treatment, four specific (and popular) models of
particle production were discussed. It was shown that the suggested
measurements provide strong constraints on all of them.

The method seems rather general and flexible. It can be applied to symmetric,
as well as to asymmetric processes. It can be used to study distributions in
various kinematic variables (e.g. rapidity and transverse momentum
\cite{bzbf}). Finally, it does not require  full acceptance of the detector.

We conclude that studies of long-range correlations in multiparticle production
may become a powerful instrument in investigations of particle production
mechanisms at high energy.

\section{Appendix }

In this appendix we calculate the number of moments of order $r$ for the
reflection symmetric case, i.e. when

\begin{equation}\label{refsym}
  F_{i_1\ldots i_B} \equiv F_{i_B\ldots i_1}.
\end{equation}
We will denote the number of these moments by $m_S(r,B)$. Let us call symmetric
the moments for which the ordered sets $\{i_1\ldots i_B\}$ and $\{i_B\ldots
i_1\}$ coincide. The number of such moments will be denoted $S(r,B)$. The
constraint (\ref{refsym}) does not affect the number of symmetric moments, but
reduces the number of the other independent moments by a factor of two.  Thus

\begin{equation}
  m_S(r,B) = \frac{1}{2}[m(r,B) + S(r,B)],
\end{equation}
where $m(r,B)$ is given by formula (\ref{nummom}) and $S(r,B)$ remains to be
calculated.

Let us begin by the case when $B = 2K+1$, where $K$ is an integer. Then the
generic form of a symmetric moment is $F_{i_1,\ldots,i_K,n,i_K,\ldots,i_1}$.
Therefore, the number of such moments can be calculated as follows. Include all
the nonnegative integers $p$ such that $n = r-2p \geq 0$. Notice that for each
$p$ there are $m(p,K)$ moments, thus in this case

\begin{equation}
  S(r,B) = \sum_p\left(\begin{array}{c}
  p+K-1 \\
  p
\end{array}\right).
\end{equation}
The case $B = 2K$ reduces to the previous one with the constraint that $n=0$.
For $r = 2I+1$, where $I$ i an integer, there are no solutions for $p$,
therefore

\begin{equation}
  S(r,B) = 0,
\end{equation}
while for $r = 2I$ the only solution is $p = I$, so that

\begin{equation}
S(r,B) = \left(\begin{array}{c}
  I+K-1 \\
  I
\end{array}\right).
\end{equation}

\section{Acknowledgements}

 This investigation was supported in part by the grant N N202
125437 of the Polish Ministry of Science and Higher Education (2009-2012).

\end{document}